\documentclass[aps,prl,twocolumn,showpacs,superscriptaddress]{revtex4}
\usepackage{graphicx}

\newlength\figurewidth
\begin{document}
\newcommand{\rem}[1]{}

\title{ Spontaneously Symmetry-Broken Current in Coupled Nanomechanical Shuttles}

\author{ Kang-Hun Ahn}
\author{Hee Chul Park}
\affiliation{ Department of Physics, Chungnam National University,
 Daejeon 305-764, Republic of Korea}

\author{ Jan Wiersig}
\affiliation{ Institut f\"ur Theoretische Physik, Universtit\"at
Bremen, Postfach 330 440, D-28334 Bremen, Germany }

\author{ Jongbae Hong}
\affiliation{ School of Physics and Astronomy, Seoul National
University, 156-747, Seoul, Korea}


\begin{abstract}
We investigate the transport and the dynamical properties of
tunnel-coupled double charge shuttles. The oscillation frequencies
of two shuttles are mode-locked to integer multiples of the applied
voltage frequency $\omega$. We show that left/right-symmetric double
shuttles may generate direct net current due to bistable motions
caused by parametric instability.
 The symmetry-broken direct current appears near $\omega =\Omega_{0}/(2j-1)$,
($j=1,2,\dots$), where $\Omega_{0}$ is the dressed resonance
frequency of the relative motion of the two shuttles.
\end{abstract}
 \maketitle

Recently, nanoelectromechanical systems (NEMS) have attracted great
attention due to fundamental aspects in new electrical transport
properties as well as new technology complementary to conventional
MEMS engineering~\cite{Cleland02}. A prototype of NEMS, single
electron shuttle, was suggested by Gorelik {\it et
al.}~\cite{Gorelik98} which is a single electron transistor combined
with its mechanical degree of freedom. The charge shuttles can be
realized in two different ways in experiments. In a top-down approach,
the charge shuttles can be realized by miniaturization of silicon
structure~\cite{Erbe01,Scheible04}. The other way is a bottom-up
approach, where the charge shuttle is produced from molecules such as
$C_{60}$~\cite{Park00}.

In this Letter, we report a theoretical study  on the transport
properties and the dynamics when two charge shuttles are coupled
through electron tunneling. The nonlinearity involved in this system
shows an interesting bistable regime, where the electric current
characteristics are of potentially great importance in NEMS
applications. An interesting situation arises when the two shuttles
are totally symmetric. In this case, through a dynamical symmetry
breaking the system produces a net direct electric current.
Symmetry-broken electric current under time-periodic perturbation
has been an interesting topic of many theories and
experiments~\cite{Mani02} where a complete understanding is still
necessary. We show that the double charge shuttle (DCS) allows for
symmetry-broken DC current caused by parametric instability. This is
in contrast to the single shuttle system where the DC current is
vanishing in symmetric configuration.

\begin{figure}
\includegraphics[width=0.8\figurewidth]{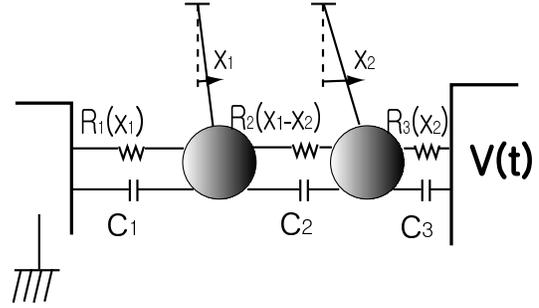}
 \caption{
A schematic figure of the double charge shuttle.} \label{fig:system}
\end{figure}

We start with the formalism of single electron
tunneling~\cite{single92} and classical dynamics which has been used
for single charge shuttle~\cite{Gorelik98,Pistolesi05}. The drain
lead is in the left side of the source lead where the ac voltage
$V(t) = V_0\sin{(\omega t)}$ is applied to the source compared to
the drain, see Fig.~\ref{fig:system}.
 The capacitance is supposed to be not sensitive to the displacement
 while
the resistance is a function of displacements $x_{1}$ and $x_{2}$:

 \begin{eqnarray}
 \nonumber
R_{1}(x_{1}) &=&R_{1}(0)e^{x_{1}/\lambda},\\
R_{2}(x_{1}-x_{2}) &=& R_{2}(0)e^{(x_{2}-x_{1})/\lambda},\\
\nonumber
 R_{3}(x_{2}) &=& R_{3}(0)e^{-x_{2}/\lambda}.
\end{eqnarray}
 Here, $\lambda$ is a phenomenological tunneling
length. When the mutual capacitance of the $j$-th junction is $c_{j}$,
the capacitance matrix is constructed as
$C_{kl}=c_{k}+c_{k+1},-c_{k+1},-c_{k},0$ for $l=k,l=k+1,l=k-1$,
otherwise, respectively. The internal charging energy is given by
$\epsilon(Q_{1},Q_{2})=\frac{1}{2}\sum^{2}_{k,l}(C^{-1})_{kl}Q_{k}Q_{l}$.
The energy loss $E_{j}$ of the $j$-th junction is
$E_{1}=\frac{c_{\mbox{\footnotesize tot}}}{c_{1}}eV+\epsilon(Q_{1},Q_{2})-\epsilon(Q_{1}-e,Q_{2}),
E_{2}=\frac{c_{\mbox{\footnotesize tot}}}{c_{2}}eV+\epsilon(Q_{1},Q_{2})-\epsilon(Q_{1}+e,Q_{2}-e),$
and
$E_{3}=\frac{c_{\mbox{\footnotesize tot}}}{c_{3}}eV+\epsilon(Q_{1},Q_{2})-\epsilon(Q_{1},Q_{2}+e)$
with the total capacitance $c_{\mbox{\footnotesize tot}}$. The rate of the
tunneling from left to the right at the $j$-th junction at low temperature and
environmental impedance is
\begin{eqnarray}
\label{trate}
\overrightarrow{\Gamma}^{(j)}_{n_{1},n_{2}}=\frac{1}{e^{2}R_{j}}E_{j}(V,Q_{1},Q_{2})\Theta[E_{j}(V,Q_{1},Q_{2})],
\end{eqnarray}
where the island charges $Q_{1},Q_{2}$ are integer multiple of
$-e$; $Q_{i}=-n_{i}e$. By replacing $V,Q_{1},Q_{2}$ with
$-V,-Q_{1},-Q_{2}$ in the right-hand side of the above equation, one
can get $\overleftarrow{\Gamma}^{(j)}_{n_{1},n_{2}}$.

Now we come to the equations of motion. The time evolution of the
probability $P_{n_{1},n_{2}}$ for the island charges
$Q_{1}=-n_{1}e$, $Q_{2}=-n_{2}e$  is given by the following rate
equation $\frac{dP_{n_{1},n_{2}}}{dt} =
\overrightarrow{\Gamma}^{(1)}_{n_{1}-1,n_{2}}P_{n_{1}-1,n_{2}} +
\overleftarrow{\Gamma}^{(1)}_{n_{1}+1,n_{2}}P_{n_{1}+1,n_{2}} +
\overrightarrow{\Gamma}^{(2)}_{n_{1}-1,n_{2}+1}P_{n_{1}-1,n_{2}+1} +
\overleftarrow{\Gamma}^{(2)}_{n_{1}+1,n_{2}-1}P_{n_{1}+1,n_{2}-1}
+\overrightarrow{\Gamma}^{(3)}_{n_{1},n_{2}-1}P_{n_{1},n_{2}-1} +
\overleftarrow{\Gamma}^{(3)}_{n_{1},n_{2}+1}P_{n_{1},n_{2}+1}
-\sum_{j=1}^3(\overleftarrow{\Gamma}^{(i)}_{n_{1},n_{2}}
+\overrightarrow{\Gamma}^{(j)}_{n_{1},n_{2}})P_{n_{1},n_{2}}$.

At low temperatures the fluctuation of the displacements $x_{1},
x_{2}$ are negligible compared to the charge fluctuations. While the
charge fluctuations are important to the noise properties, we
suppose $1/\Gamma $ be much smaller than the typical time scale of
displacement and investigate the mechanical motions using the mean
island charge $\langle Q_{i}(t)\rangle=-e\langle
n_{i}(t)\rangle=-e\sum_{n_{1},n_{2}}n_{i}P(n_{1},n_{2},t)$. The
island charges experience the force produced by the electric field
$-V(t)/L$;
\begin{eqnarray}
\ddot{x_{i}}+\gamma_i\dot{x_{i}}+\omega^{2}_{0i}x_{i}=-\frac{V(t)}{m_i
L}\langle Q_{i}(t)\rangle,~~~~i=1,2 \label{eom}
\end{eqnarray}
where $L$ is the source-drain distance and $\gamma_{i}$,
$\omega_{0i}$ denote the friction constant and the natural angular
frequency of $i$-th shuttle. Note that the resonators are modelled as linear
oscillators. The nonlinearity of the full system comes from the coupling via
tunneling. The
electric current from the source in
the right-hand side to the drain in the left-hand side is computed
as
\begin{eqnarray}\label{eq:current}
I(t)=e\sum_{n_{1},n_{2}}(
\overrightarrow{\Gamma}^{(1)}_{n_{1},n_{2}}
-\overleftarrow{\Gamma}^{(1)}_{n_{1},n_{2}} )P_{n_{1},n_{2}}(t) \ .
\end{eqnarray}

\begin{figure}
\includegraphics[width=0.8\figurewidth]{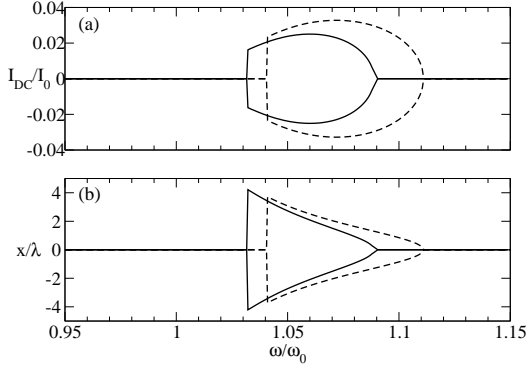}
\caption{(a) The symmetry-broken DC current in units of
$I_{0}=V_{0}/R$ vs. frequency of the applied AC
  voltage. The solid line refers to the result from the full
  calculations mentioned in the text and the dashed line refers to
  the result from the simplified calculation in the adiabatic limit.  The
 parameters are fixed to $c =100\,$aF, $R= h/e^2$,
$\lambda = 20\,$nm, $L = 500\,$nm, $\omega_{0} = 50\,$MHz, $\gamma =
0.025\,\omega_{0}$, $m = 1.1\times10^{-21}\,$Kg, and $V_0 =
10.8\,$mV. (b) The corresponding stroboscopic plot of $
x=x_{1}-x_{2}$
  obtained for a fixed phase of the oscillating voltage.}
\label{fig:po-curr}
\end{figure}

In the following we will consider a symmetric configuration: $\omega_{0i} =
\omega_0$,
$\gamma_i = \gamma$, $m_i = m$, $R_{1}(0)=R_{3}(0) = 0.5R_{2}(0) \equiv R$, and
$c_{1}=c_{3}= 2c_{2} \equiv c$.
In Fig.~\ref{fig:po-curr}(a), we plot the time-averaged current
$I_{\mbox{\footnotesize
DC}}=\frac{\omega}{2\pi}\int_{0}^{2\pi/\omega}I(t)dt$. From the
symmetry of the configuration one would expect that the DC current
computed according to Eq.~(\ref{eq:current}) is zero for all
frequencies of the applied AC voltage. Fig.~\ref{fig:po-curr}(a),
however, shows that in the vicinity of the natural frequency of the
oscillators, $\omega_0$, the DC current can be negative or positive
depending on the initial conditions. The initial conditions are
specified by the initial position $x_1$, $x_2$ and the velocities
$\dot{x}_1$, $\dot{x}_2$ at a given starting time. In the regime of
zero current all initial conditions are attracted to an unique
stable asymptotic solution (the so-called attractor). In the nonzero
current case, two such attractors coexist and cause a {\it dynamical
symmetry breaking}.

The global behaviour of the attractors can be conveniently
represented on a stroboscopic section (a special case of a
Poincar\'e section~\cite{AP90}) defined by a fixed value of the
phase of the driving voltage. Fig.~\ref{fig:po-curr}(b) shows the
stroboscopic section at a fixed phase of the applied voltage. We
find the center of mass coordinate does not move; $X=(x_{1}+x_{2})/2
\approx 0$ for all frequencies $\omega$. Meanwhile in the regime of
nonzero DC current we find two symmetry-related solutions for the
relative coordinate $x=x_1-x_2$.

We may get further insight by investigating the adiabatic limit of
the system. In Fig.~\ref{fig:po-curr} we plot the DC current and relative
coordinates computed from the adiabatic approach using dotted lines.
One can notice that the full numerical results are quite close to
those obtained in the adiabatic limit. Therefore we may rely on the
adiabatic approach for the analysis of the bistability.

In the adiabatic limit the electronic relaxation is much faster than the
mechanical motion, i.e.
\begin{eqnarray}
\omega R c \;, \;\omega_{0} R c \ll 1.
\end{eqnarray}
 In this case, a classical circuit analysis of Fig.~\ref{fig:system} on the
charges $q_{j}$ accumulated in each capacitor $c_{j}$ gives
\begin{eqnarray}
\nonumber \frac{q_{1}}{R_{1}c_{1}}-\frac{q_{2}}{R_{2}c_{2}}=
\frac{q_{2}}{R_{2}c_{2}}-\frac{q_{3}}{R_{3}c_{3}}&=&0\\
\frac{q_{1}}{c_{1}}+\frac{q_{2}}{c_{2}}+\frac{q_{3}}{c_{3}}&=&V(t) \ .
\end{eqnarray}
From the solutions of the above
linear equations, the net charges of each islands
$Q_{1}=q_{1}-q_{2}$, $Q_{2}=q_{2}-q_{3}$ are simply given by
\begin{eqnarray}
Q_1 & = &
Vc\frac{e^{x_1/\lambda}-e^{(x_2-x_1)/\lambda}}{e^{x_1/\lambda}+2e^{(x_2-x_1)/\lambda}
+e^{-x_2/\lambda}} \ ,\\
Q_2 & = &
Vc\frac{e^{(x_2-x_1)/\lambda}-e^{x_2/\lambda}}{e^{x_1/\lambda}+2e^{(x_2-x_1)/\lambda}
+e^{-x_2/\lambda}} \ .
\end{eqnarray}

Plugging these into the equation of motion for the mechanical degree
of freedom~(\ref{eom}) gives the equation of motion for the center
of mass coordinate
\begin{equation}
\ddot{X}+\gamma\dot{X}+\omega_0^2X =
\left(-\frac{cV(t)^2}{2mL}\right)
\frac{e^{X/\lambda}-e^{-X/\lambda}}{e^{X/\lambda}+2e^{-3
x/2\lambda}+e^{-X/\lambda}},
\end{equation}
where $x=x_{1}-x_{2}$ is the relative coordinate. Clearly, $X=0$ is
a solution, in agreement with our numerical finding that the center
of mass is not moving. Exploiting this fact, the equation of motion
for the relative coordinate $x$ can be derived as
\begin{equation}\label{eq:tanhx}
\ddot{ x}+\gamma\dot{x}+\omega_0^2 x = -\frac{cV_{0}^2\sin^{2}\omega
t}{mL} \tanh{\frac{3 x}{4\lambda}} \ .
\end{equation}

From the above equation one can see that $x=0$ is a trivial
solution. However, this solution can be unstable as we will prove in
the following. Note that if a nontrivial solution of $x(t)$ exists
then $-x(t)$ is also a solution. This is the pair of bistable
solutions.  Since the equation is invariant under the
time-translation operation $t \rightarrow t+\pi/\omega$, a periodic
solution should satisfy $x(t+\pi/\omega)=\pm x (t)$. As will be
shown later, this parity is important for the nonzero DC current.

The electric current in the adiabatic limit can be expressed as
$I(t)=q_{1}(t)/[R_{1}(x)c_{1}]=V(t)/[R_{1}(x)+R_{2}(x_{1},x_{2})+R_{3}(x_{2})]$.
Therefore the time-averaged DC current reads
\begin{eqnarray}
\label{current}
 I_{\mbox{\footnotesize DC}}=I_{0}\frac{\omega }{4\pi}\int_{t_{0}}^{t_{0}+2\pi/\omega} \frac{\sin \omega
t}{e^{x(t)/2\lambda}+e^{-x(t)/\lambda}} dt \ ,
\end{eqnarray}
where $I_{0}=V_{0}/R$.  The dashed line in Fig.~\ref{fig:po-curr}(a)
was obtained using the above formula which also shows a clear
bistability.

 Now we are going to show
that the origin of the symmetry-broken current and the bistability
is parametric instability. By linearizing the term
$\tanh{(3x/4\lambda)}$ in the right hand side of
Eq.~(\ref{eq:tanhx}), we get
\begin{equation}\label{eq:Mathieu}
\ddot{ x}+\gamma\dot{ x}+\Omega_0^2\left[1-\frac{\mu^{2}
}{1+\mu^{2}}\cos{(2\omega
    t)}\right] x = 0,
\end{equation}
where $\mu=\frac{V_{0}}{\omega_{0}}\sqrt{\frac{3c}{8\lambda m L }}$
and $\Omega_{0}$ is the dressed harmonic frequency of the relative
motion;
\begin{eqnarray}
\Omega_{0}=\omega_{0}\sqrt{1+\mu^{2}}=\sqrt{\omega_{0}^{2}+\frac{3cV_{0}^{2}}{8\lambda
m L } } \ .
\end{eqnarray}
Equation~(\ref{eq:Mathieu}) is called a damped Mathieu equation which is
a paradigm for studying parametric resonance~\cite{Butikov04}. The
stability analysis in Ref.~\cite{Butikov04} up to second order of
$M$ shows that the motion is unstable in the interval $\omega_{-} <
\omega < \omega_{+}$ where
\begin{eqnarray}
\label{analytic}
 \omega_{\pm}=\Omega_{0}\left(1\pm
\frac{1}{2}\sqrt{M^{2}-\gamma^{2}/\Omega_{0}^{2}}+\frac{11}{16}M^{2}\right),
\end{eqnarray}
with $M=\frac{\mu^{2}}{2(1+\mu^{2})}$.
 This interval is called principal instability interval. Obviously, the
interval has finite width only if $M>\gamma/\Omega_{0}$, i.e. the
strength of the driving, $V_0$, must be sufficiently large in order
to get parametric instability.

\begin{figure}
\includegraphics[width=0.8\figurewidth]{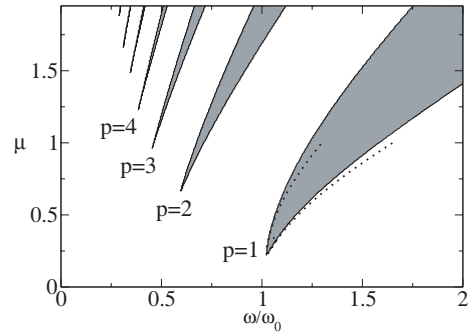}
\caption{Bifurcation diagram $\mu$ vs. $\omega/\omega_0$ in the
adiabatic limit for $\gamma=0.025\omega_{0}$. Bistable regions
($Arnol'$d tongues) are mode-locked with $p = 1,2,3,4,\dots$. The
nonzero DC current arises only when $p$ is odd. The dotted line
denotes the boundary of the principal instability region obtained
from the analytic formula~(\ref{analytic}).}
 \label{fig:tongues}
\end{figure}

From the above analysis and Fig.~\ref{fig:po-curr}(b) we conclude that when
entering the principal instability interval from the right hand side the
unique attractor turns into an unstable solution (a repeller) thereby creating
two new attractors.
This scenario leading to bistability is termed a supercritical pitchfork
bifurcation. The bifurcation occuring when entering the principal
instability interval from the left hand side is different which can be seen
from the abrupt change of the value of $x$. Here we have three attractors and
two repellers (not shown). The repellers and the central attractor ($x=0$)
merge in a subcritical pitchfork bifurcation creating a repeller at $x=0$.

In Fig.~\ref{fig:tongues}, we plot the phase diagram of $\mu$ and
$\omega/\omega_{0}$ which indicates the locations of the bistable
regimes computed from Eq.~(\ref{eq:tanhx}). The dotted line denote
the analytic results in Eq.~(\ref{analytic}) from the linearized
equation (\ref{eq:Mathieu}). The reasonable agreement of the
parameters showing the instability confirms that the parametric
amplification causes the instability in the system. Beside the
principal instability region there are higher-order regions in which
the ratio of the frequency of the oscillators and the applied
voltage is mode-locked to $p = 2,3,\ldots$. These regions, and also
the principal instability region, are called {\it Arnol$'$d
tongues}~\cite{Arnold83}.

\begin{figure}
\includegraphics[width=0.8\figurewidth]{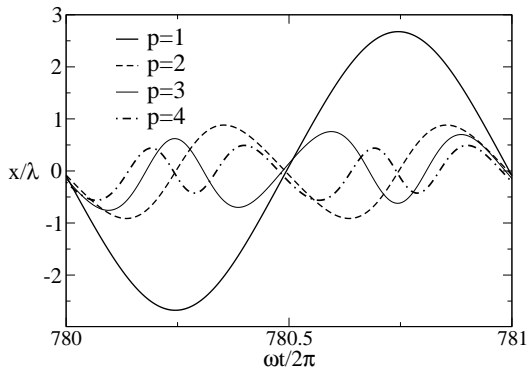}
\caption{Time evolution of the relative coordinate $x=x_{1}-x_{2}$
for typical sets of values ($\omega/\omega_{0},\mu$) belonging to
the bistable regimes of $p=1,2,3,4$;
$\omega/\omega_{0}=1.45,0.79,0.53,0.415$ and $\mu=1.414$.}
 \label{fig:pos-time}
\end{figure}

The second interval of instability denoted by $p=2$ in
Fig.~\ref{fig:tongues}
arise from second harmonics of the system. The analytic formula for
this region from the stationary solutions in Ref.~\cite{Butikov04}
gives
\begin{eqnarray}
\omega_{\pm}=\frac{1}{2}\Omega_{0}\left(1+\frac{5}{12}M^{2}\pm\frac{1}{8}\sqrt{M^{4}-(4\gamma/\Omega_{0})^{2}}\right).
\end{eqnarray}
The higher order bistable regions are found in our numerical
calculations. The center of the intervals in each bistable regions
are located where
\begin{eqnarray}
\omega \approx \frac{\Omega_{0}}{p}\,;~~~~p=1,2,3,\dots .
\end{eqnarray}
The main characteristics of the each bistable region is that the
(unharmonic) frequency of the relative motion is given by $p\omega$
 even though there is a mismatch between the natural harmonic frequency and
the frequency of the voltage, which is called {\it
  mode-locking} in the field of nonlinear dynamics~\cite{AP90}.
In Fig.~\ref{fig:pos-time}, we plot the time series of the relative coordinate
$x$ for
various frequencies corresponding to different bistable regimes. One can
clearly see the $p$-th order mode-locking. For instance, in spite of the
large mismatch between
$\omega$ and $\omega_{0}$ when $p=1$  ($\omega=1.45\omega_{0}$) the
oscillation frequency is clearly given by~$\omega$.

It is important to note that the rectified DC current arises only
when $p$ is an odd number of integer. When $p$ is even, the periodic solution
$x(t)$ has even parity i.e.
$x(t+\pi/\omega)=x(t)$.  In this case the integration in
Eq.~(\ref{current}) from $t_{0}$ to $t_{0}+\pi/\omega$ is cancelled by
the integration from $t_{0}+\pi/\omega$ to $t_{0}+2\pi/\omega$.

One may ask whether the parametric instability can cause symmetry
breaking already in a single shuttle system.  When we write the
tunnel resistances as $R_{1}(x)=R(0)e^{+x/\lambda}$ and
$R_{2}(x)=R(0)e^{- x/\lambda}$ where $x$ is the coordinate of the
shuttle, the equation of motion is given by $\ddot{
x}+\gamma\dot{x}+\omega_0^2 x = -\frac{cV_{0}^2\sin^{2}\omega t}{mL}
\tanh{\frac{ x}{\lambda}} $ similar to Eq.~(\ref{eq:tanhx}).
Therefore,in the sense of the mechanical motion, the parametric
resonance gives rise to the bistability and the bifurcation diagram
is quite similar to Fig.~\ref{fig:tongues}. However, {\it the single
shuttle systems does not allow a symmetry-broken DC current}. The
absence of the DC current in the symmetric single shuttle is clear
from the time-averaged current formula of the system
\begin{eqnarray}
I_{\mbox{\footnotesize DC}} =I_{0}\frac{\omega}{4\pi}\int_{t_{0}}^{t_{0}+2\pi/\omega}\frac{\sin\omega
t}{\cosh x/\lambda} dt.
\end{eqnarray}
Note that the symmetry of the periodic solution
$x(t+\pi/\omega)=\pm x(t)$ always ensures
$\cosh{[x(t+\pi/\omega)/\lambda]}=\cosh{[x(t)/\lambda]}$ confirming that the
above integral is zero.

In experimental systems, the charge and displacement fluctuations
exist due to the discreetness of charges and finiteness of
temperature. Noise properties of two colloidal particles in DC
source-drain bias have been numerically investigated in
Ref.\cite{Nishiguchi02} where the shot noise produces random
telegraph noise. The noise properties in the presence of oscillating
source-drain voltage are expected to contain interesting information
on the mechanical properties, which is currently under study. For
instance, the shuttles are expected to show enhanced noise power in
the bistable regimes we studied in this work.

Our results are robust against the finite temperature effect which
modifies the tunneling rate in Eq.(\ref{trate}).This is because the
dynamical symmetry breaking appears as a classical effect. While the
perfect left/right symmetry does not exist when the tunnel
resistances differ by multiple of 5 or 10, we find the effects are
still visible from the enhancement of the net current. The enhanced
net current is also visible when the resonance frequencies and
tunneling coefficients differ by 10 percents. This robustness comes
from the nonlinear phenomenon, {\it mode-locking}, where the
mechanical motions are locked to the driving voltage in spite of
small variations of the system parameters. One may also note that
the tunneling length $\lambda$ enters the instability condition
through $\mu$,  so the net current can exist in a reasonable range
of $\lambda$ if $\mu$ and $\omega$ belongs to the Arnol$'$d tongue.

In summary, the transport through two tunnel-coupled symmetric
charge shuttles has been studied. We found that the oscillation frequencies of
the two shuttles are mode-locked to the applied oscillating
voltage. Moreover, we observed a dynamical bistability which allows for
non-zero electric currents. The origin of this phenomenon has been traced back
to parametric instability induced by the nonlinear coupling of the
mechanical and electrical degree of freedom.

\begin{acknowledgements}
We thank Daniel Park, Raj Mohanty, Andrew Cleland, Boris Altshuler,
Robert Shekter, and Fabio Pistolesi for useful and encouraging
discussions. This work was supported by the Korea Research
Foundation (Grant Nr. KRF-2006-311-C00059).
\end{acknowledgements}


\begin{thebibliography}{12}
\expandafter\ifx\csname natexlab\endcsname\relax\def\natexlab#1{#1}\fi
\expandafter\ifx\csname bibnamefont\endcsname\relax
  \def\bibnamefont#1{#1}\fi
\expandafter\ifx\csname bibfnamefont\endcsname\relax
  \def\bibfnamefont#1{#1}\fi
\expandafter\ifx\csname citenamefont\endcsname\relax
  \def\citenamefont#1{#1}\fi
\expandafter\ifx\csname url\endcsname\relax
  \def\url#1{\texttt{#1}}\fi
\expandafter\ifx\csname urlprefix\endcsname\relax\def\urlprefix{URL }\fi
\providecommand{\bibinfo}[2]{#2}
\providecommand{\eprint}[2][]{\url{#2}}

\bibitem[{\citenamefont{Cleland}(2002)}]{Cleland02}
\bibinfo{author}{\bibfnamefont{A.}~\bibnamefont{Cleland}},
  \emph{\bibinfo{title}{Foundation of Nanomechanics}}
  (\bibinfo{publisher}{Springer}, \bibinfo{address}{Heidelberg},
  \bibinfo{year}{2002}).

\bibitem[{\citenamefont{Gorelik et~al.}(1998)\citenamefont{Gorelik, Isacsson,
  Voinova, Kasemo, Shekhter, and Jonson}}]{Gorelik98}
\bibinfo{author}{\bibfnamefont{L.~Y.} \bibnamefont{Gorelik}},
  \bibinfo{author}{\bibfnamefont{A.}~\bibnamefont{Isacsson}},
  \bibinfo{author}{\bibfnamefont{M.~V.} \bibnamefont{Voinova}},
  \bibinfo{author}{\bibfnamefont{B.}~\bibnamefont{Kasemo}},
  \bibinfo{author}{\bibfnamefont{R.~I.} \bibnamefont{Shekhter}},
  \bibnamefont{and} \bibinfo{author}{\bibfnamefont{M.}~\bibnamefont{Jonson}},
  \bibinfo{journal}{Phys. Rev. Lett.} \textbf{\bibinfo{volume}{80}},
  \bibinfo{pages}{4526} (\bibinfo{year}{1998}).

\bibitem[{\citenamefont{Erbe et~al.}(2001)\citenamefont{Erbe, Weiss, Zwerger,
  and Blick}}]{Erbe01}
\bibinfo{author}{\bibfnamefont{A.}~\bibnamefont{Erbe}},
  \bibinfo{author}{\bibfnamefont{C.}~\bibnamefont{Weiss}},
  \bibinfo{author}{\bibfnamefont{W.}~\bibnamefont{Zwerger}}, \bibnamefont{and}
  \bibinfo{author}{\bibfnamefont{R.~H.} \bibnamefont{Blick}},
  \bibinfo{journal}{Phys. Rev. Lett.} \textbf{\bibinfo{volume}{87}},
  \bibinfo{pages}{096106} (\bibinfo{year}{2001}).

\bibitem[{\citenamefont{Scheible and Blick}(2004)}]{Scheible04}
\bibinfo{author}{\bibfnamefont{D.~V.} \bibnamefont{Scheible}} \bibnamefont{and}
  \bibinfo{author}{\bibfnamefont{R.~H.} \bibnamefont{Blick}},
  \bibinfo{journal}{Appl. Phys. Lett.} \textbf{\bibinfo{volume}{84}},
  \bibinfo{pages}{4632} (\bibinfo{year}{2004}).

\bibitem[{\citenamefont{Park et~al.}(2000)\citenamefont{Park, Park, Lim,
  Anderson, Alivisatos, and McEuen}}]{Park00}
\bibinfo{author}{\bibfnamefont{H.}~\bibnamefont{Park}},
  \bibinfo{author}{\bibfnamefont{J.}~\bibnamefont{Park}},
  \bibinfo{author}{\bibfnamefont{A.~K.~L.} \bibnamefont{Lim}},
  \bibinfo{author}{\bibfnamefont{E.~H.} \bibnamefont{Anderson}},
  \bibinfo{author}{\bibfnamefont{A.~P.} \bibnamefont{Alivisatos}},
  \bibnamefont{and} \bibinfo{author}{\bibfnamefont{P.~L.}
  \bibnamefont{McEuen}}, \bibinfo{journal}{Nature}
  \textbf{\bibinfo{volume}{407}}, \bibinfo{pages}{57} (\bibinfo{year}{2000}).

\bibitem[{\citenamefont{Mani et~al.}(2002)\citenamefont{Mani, Smet, von
  Klitzing, Narayanamurti, Johnson, and Umansky}}]{Mani02}
\bibinfo{author}{\bibfnamefont{R.~G.} \bibnamefont{Mani}},
  \bibinfo{author}{\bibfnamefont{J.~H.} \bibnamefont{Smet}},
  \bibinfo{author}{\bibfnamefont{K.}~\bibnamefont{von Klitzing}},
  \bibinfo{author}{\bibfnamefont{V.}~\bibnamefont{Narayanamurti}},
  \bibinfo{author}{\bibfnamefont{W.~B.} \bibnamefont{Johnson}},
  \bibnamefont{and} \bibinfo{author}{\bibfnamefont{V.}~\bibnamefont{Umansky}},
  \bibinfo{journal}{Nature} \textbf{\bibinfo{volume}{420}},
  \bibinfo{pages}{646} (\bibinfo{year}{2002}).

\bibitem[{\citenamefont{Grabert and Devoret}(1992)}]{single92}
\bibinfo{author}{\bibfnamefont{H.}~\bibnamefont{Grabert}} \bibnamefont{and}
  \bibinfo{author}{\bibfnamefont{M.~H.} \bibnamefont{Devoret}},
  \emph{\bibinfo{title}{Single Charge Tunneling-Coulomb Blockade Phenomena in
  Nanostructures}} (\bibinfo{publisher}{Plenum}, \bibinfo{address}{New York},
  \bibinfo{year}{1992}).

\bibitem[{\citenamefont{Pistolesi and Fazio}(2005)}]{Pistolesi05}
\bibinfo{author}{\bibfnamefont{F.}~\bibnamefont{Pistolesi}} \bibnamefont{and}
  \bibinfo{author}{\bibfnamefont{R.}~\bibnamefont{Fazio}},
  \bibinfo{journal}{Phys. Rev. Lett.} \textbf{\bibinfo{volume}{94}},
  \bibinfo{pages}{036806} (\bibinfo{year}{2005}).

\bibitem[{\citenamefont{Arrowsmith and Place}(1990)}]{AP90}
\bibinfo{author}{\bibfnamefont{D.~K.} \bibnamefont{Arrowsmith}}
  \bibnamefont{and} \bibinfo{author}{\bibfnamefont{C.~M.} \bibnamefont{Place}},
  \emph{\bibinfo{title}{An introduction to dynamical systems}}
  (\bibinfo{publisher}{Cambridge University Press},
  \bibinfo{address}{Cambridge}, \bibinfo{year}{1990}).

\bibitem[{\citenamefont{Butikov}(2004)}]{Butikov04}
\bibinfo{author}{\bibfnamefont{E.~I.} \bibnamefont{Butikov}},
  \bibinfo{journal}{Eur. J. Phys.} \textbf{\bibinfo{volume}{25}},
  \bibinfo{pages}{535} (\bibinfo{year}{2004}).

\bibitem[{\citenamefont{Arnol$'$d}(1983)}]{Arnold83}
\bibinfo{author}{\bibfnamefont{V.~I.} \bibnamefont{Arnol$'$d}},
  \bibinfo{journal}{Russian Mathematical Surveys}
  \textbf{\bibinfo{volume}{38}}, \bibinfo{pages}{215} (\bibinfo{year}{1983}).

\bibitem[{\citenamefont{Nishiguchi02}(2002)}]{Nishiguchi02}
\bibinfo{author}{\bibfnamefont{N.} \bibnamefont{Nishiguchi}},
 \bibinfo{journal}{Phys. Rev. Lett.}
  \textbf{\bibinfo{volume}{89}}, \bibinfo{pages}{066802} (\bibinfo{year}{2002}).



\end{thebibliography}

\end{document}